\documentstyle[prd,floats,aps]{revtex}
\input epsf
\begin{document}
\title{The collision of boosted black holes}
\author{John Baker$^1$,
Andrew Abrahams $^2$,
Peter Anninos$^2$,
Steve Brandt$^{2,3,4}$,\\
Richard Price$^5$,
Jorge Pullin$^1$ and
Edward Seidel$^{2,3,4,6}$\\
1. {\it Center for Gravitational Physics and Geometry, Department of
Physics\\The Pennsylvania State University, 104 Davey Lab, University
Park, PA 16802}\\
2. {\it National Center for Supercomputing Applications,
605 E. Springfield Ave., Champaign, IL 61820}\\
3. {\it Max-Plank-Institute f\"ur Gravitationphysik,
Albert-Einstein-Institute, 14473 Potsdam, Germany}\\
4. {\it Department of Physics, University of Illinois at
Urbana-Champaign, 61801}\\
5. {\it Department of Physics, University of Utah, Salt Lake City, UT
84112}\\
6.  {\it Department of Astronomy, University of Illinois at
Urbana-Champaign, 61801}}

\maketitle
\begin{abstract}
  We study the radiation from a collision of black holes with equal
  and opposite linear momenta. Results are presented from a full
  numerical relativity treatment and are compared with the results
  from a ``close-slow'' approximation. The agreement is remarkable,
  and suggests several insights about the generation of gravitational
  radiation in black hole collisions.
\end{abstract}

\vspace{-7.5cm} 
\begin{flushright}
\baselineskip=15pt
CGPG-96/8-3  \\
gr-qc/9608064\\
\end{flushright}
\vspace{5.5cm}
\section{Introduction}

The collision of two black holes is now being studied extensively via
the techniques of numerical relativity
\cite{grandchallenge}. Collisions are of great importance as the most
interesting source of gravitational waves that might be observable
with interferometric detectors\cite{ligo}. The study is also of great
inherent interest to relativity theory in that supercomputers allow us
to investigate strong field gravity effects without symmetries which
might preclude interesting or crucial phenomena.  In dealing with such
a daunting problem, useful checks, guidelines, and insights have been
provided by analytical approximations, in particular by the
close-limit approximation \cite{AbPr}.  In principle, this method
applies when the holes are initially very close together. In this
case, the horizon is initially only slightly nonspherical and the
spacetime that evolves outside the horizon can be treated as a
perturbed single black hole. The highly nonspherical nature of the
spacetime inside the horizon is causally disconnected from the
exterior, and from the generation of outgoing gravitational waves.
The exterior spacetime can be evolved forward in time from the initial
data hypersurface with the linearized equations of perturbation
theory.

This method turns out to be remarkably
successful\cite{PrPu,Anninosetal,AbPr2}. The details of this success
may give insights into the nature of collisions of holes.  For holes
that are initially momentarily stationary, the close-limit predictions
of radiated energy and waveforms are quite good (i.e., in agreement
with the results of numerical relativity) even when the initial
horizon is highly distorted, violating the assumptions underlying the
method.  The close limit has been used by Abrahams and
Cook\cite{ac94} for  the head-on collision of holes
with initial momenta towards each other.  This momentum
causes horizons to form when the holes are at larger separation and
makes the exterior spacetime more spherical, so it is not surprising
that the close limit should be successful for these cases.  Puzzling
results emerge, however, when close-limit calculations are combined
with Newtonian trajectories to estimate the radiated
energy for initially {\it large} separations of initially stationary
holes. The success of these estimates suggests, among other things, that
to a large extent the role of the early weak-field phase of the
evolution is to only to determine what the momentum of the holes will be when
they start to interact nonlinearly.

With that suggestion as one of our motivations, we consider here equal
mass holes which are initially moving towards each other with equal
and opposite momentum $P$.  We analyze the problem with an
approximation simple enough to allow insight, and we present, for
comparison, the results of full numerical relativity for the same
initial black hole configuration.  In a certain sense, this study
complements that of  Ref. \cite{ac94}.  The
initial data sets being studied are representations of the same
physical system; in Ref.  \cite{ac94} the data were
``exact'' (up to numerical error) solutions to the initial value
problem, however, in the current study we have more control over the
approximations implicit in the perturbative analysis.  In Sec.~II we
present the general formalism for the problem and briefly discuss the
full numerical solution. In Sec.~III we describe an approximation
based on the close limit {\it and} on slow initial motion.  Results of
both methods are presented and discussed in section IV. Throughout the
paper we use units in which $c=G=1$, and $M$ represents the total ADM
mass on the initial hypersurface.

\section{Initially moving holes}

The initial value equations
for general relativity are \cite{BoYo},
\begin{eqnarray}
\nabla^a (K_{ab} - g_{ab} K) &=& 0\\
{}^3R-K_{ab} K^{ab} + K^2 &=&0
\end{eqnarray}
where $g_{ab}$ is the spatial metric, $K_{ab}$ is the extrinsic
curvature and ${}^3R$ is the scalar curvature of the three metric. One
proposes a three metric that is conformally flat $g_{ab} = \phi^4
\widehat{g}_{ab}$, with $\widehat{g}_{ab}$ a flat metric, and
$\phi^4$ the conformal factor, and one uses a decomposition of the
extrinsic curvature $K_{ab} = \phi^{-2} \widehat{K}_{ab}$. The
constraints become,
\begin{eqnarray}
\widehat{\nabla}^a \widehat{K}_{ab} &=& 0\label{momentum}\\
\widehat{\nabla}^2 \phi &=& -\frac{1}{8}
\phi^{-7} \widehat{K}_{ab} \widehat{K}^{ab}\ ,\label{hamil}
\end{eqnarray}
where $\widehat{\nabla}$ is a flat-space covariant derivative.

%%%%%%%%%%%%%%%%%%%%%%%%%%%%%%%%%%%%%%

In describing how (\ref{momentum}) and (\ref{hamil}) and the
3+1 evolution equations are solved
numerically, it is useful to have at hand three different coordinate
systems. Of greatest relevance to the numerical method are the
\v{C}ade\v{z} coordinates, a system which is particularly well-suited
for the collision of two black holes and which has been used
extensively in numerical studies\cite{Smarr,anninosPRL}. These
coordinates are spherical near the throats of both holes and in the
asymptotic wave zone, so they simplify the application of
both inner and outer boundary conditions.  It is useful
also to refer to two coordinatizations of the flat conformal three
space: cylindrical coordinates $\rho,z,\varphi$, and the
bispherical-like Misner\cite{misner} coordinates $\mu,\eta,\varphi$.
The fact that the problem is axisymmetric, of course, reduces the
spatial computational grid to a two dimensional one.
By choosing to consider only equal mass holes with equal and opposite
momenta, we have a further symmetry which reduces the size of the
computational grid to a quadrant, ($\varphi=0, z>0$). We characterize the
separation of the holes with the Misner parameter $\mu_0$, and
construct the coordinate grid independently for each choice of
$\mu_0$.  Details of the grid computation are given in Refs.
\cite{Cadez71,Anninos94a}.

To solve the momentum constraint (\ref{momentum}) we follow the
prescription of York and coworkers\cite{Kulkarni83} and
Cook~\cite{Cook90,Cook93}. This starts with a solution to  (\ref{momentum})
that represents the momentum of one hole,
\begin{equation}\label{onehole}
\hat{K}^{\rm one}_{ab} = {3 \over  2 r^2} \left[ 2 P_{(a} n_{b)} -(\delta_{ab}
-n_a n_b)P^c n_c\right]\ .  \label{boyok}
\end{equation}
Here the hole is associated with some point in
the flat conformal space, $\vec{r}$ is the vector from that point, and
$\vec{n}$ is the unit vector in the $\vec{r}$ direction.  The next
step is to modify (\ref{onehole}) to represent holes centered at
$z=\pm \coth\mu_0$, the centers of the circles $\mu=\pm\mu_0$ in the
conformally flat metric.  Since the momentum constraint
(\ref{momentum}) is linear, one can simply add two expressions of the
form (\ref{onehole}):
\begin{equation}\label{two}
\widehat{K}^{\rm two}_{ij} =
\widehat{K}^{\rm one}_{ij}\left(z \rightarrow z-\coth\mu_0 \right)+
\widehat{K}^{\rm one}_{ij}\left(z \rightarrow z+
\coth\mu_0,P \rightarrow -P\right)\ .
\end{equation}

For convenience, the initial data is forced to obey an isometry condition,i.e.,
we operate on the momentum constraint solution with a reflection
procedure equivalent to adding image charges in electrostatics. The
result of this procedure is to create a solution which corresponds to
two identical asymptotically flat universes connected by two
Einstein-Rosen bridges. The nature of this symmetrization process, and
the boundary condition it provides for (\ref{hamil}), affects the
mass of the holes being represented.  Cook\cite{Cook93,cook91} has also used
this approach to develop codes to compute symmetric initial data
solutions for axisymmetric and full 3D data.

The Hamiltonian constraint is solved by linearizing equation
(\ref{hamil}) around a solution $\phi_1$ so that $\phi = \phi_1 +
\delta \phi$, discretizing to second order the resulting linear
elliptic equation, solving the matrix equation for $\delta \phi$ with a
multigrid method, then iterating the procedure until a convergence
tolerance of $\delta\phi/\phi_1 < 10^{-10}$ is achieved.  It has been
verified that for $\widehat{K}_{ab}=0$, the solution for $\phi$
converges quadratically with cell size to the time--symmetric Misner
data\cite{misner}.

The symmetrized initial data for $\phi$ and for $K_{ab}$ are now used
as the starting point for numerical integration.
The evolution employs maximal time slices and the shift
is determined by an elliptic condition that forces the
3-metric (in Cadez coordinates) into diagonal form\cite{Anninosetal}.  The
numerical errors inherent in the method (to be described elsewhere) are
similar to those in Ref.~\cite{Anninosetal}. We have verified that the
convergence rate for the total radiated energy scales quadratically with
grid spacing and that differences in the dominant $\ell=2$ waveforms
are on the order of a few percent at the grid resolutions used here.
The errors are small on the
scale of Fig.~\ref{energy}, and do not affect any conclusions to be
drawn from that figure.  The methods used for the numerical evolution
are described in detail in Ref.\cite{anninosPRL}; we modified only slightly
the code described there for evolving the time symmetric Misner data.

%%%%%%%%%%%%%%%%%%%%%%%%%%%%%%%%%%%
\section{Approximation method}
The close-limit approach can be applied to the Cook\cite{cook91}
initial data, as has been done in Ref.
\cite{ac94}. But the Cook initial solution is
numerical. To facilitate insights we make a further approximation. We
assume that the black holes are initially close, and that the initial
momentum $P$ is small.  Our solution for the extrinsic curvature
$\widehat{K}_{ab}$ is $\widehat{K}^{\rm two}_{ab}$ from (\ref{two}),
the simple superposition (without symmetrization; this effect will be
discussed later) of two one-hole solutions. We denote by $\vec{n}^+$
and $\vec{n}^-$ the normal vectors corresponding, respectively, to the
one hole solutions at $z=+L/2$ and at $-L/2$, and we define $R$ to be
the distance to a field point, in the flat conformal space, from the
point midway between the holes.  For large $R$, the normal vectors
$\vec{n}^+$ and $\vec{n}^-$ almost cancel\cite{PuCa}.  More
specifically $\vec{n}^+=-\vec{n}^-+O(L/R)$. A consequence of this is
that the total initial $\widehat{K}^{ab}$ is first order in $L/R$, and
its ($R,\theta,\varphi$ coordinate basis) components can be written as
\begin{equation}\label{Kapprox}
\hat{K}_{ab} = {3 P L \over 2R^3} \left[
\begin{array}{ccc} -4 \cos^2 \theta&0&0\\0&R^2 (1+\cos^2 \theta)&0\\
0&0&R^2 \sin^2 \theta (3 \cos^2 \theta -1)
\end{array}\right]\ .
\end{equation}

In addition to being first order in $L$, the solution for
$\widehat{K}^{ab}$ is first-order in $P$ and therefore the source term
on the right in the hamiltonian constraint (\ref{hamil}) is quadratic
in $P$. If we limit ourselves to a solution to first order in $P$ we
can ignore this quadratic source term.  (In Sec.~IV, a more thorough
discussion will be given for this step of ignoring the source term.)
Without the source term the hamiltonian constraint reduces to the zero
momentum case, the Laplace equation. The symmetric solution to this
(i.e., the solution for two identical asymptotically flat universes)
is the Misner solution\cite{misner}, and this is the solution we take.
The Misner geometry is characterized by a dimensionless parameter
$\mu_0$ which describes the separation of the throats.  We must, of
course, choose $\mu_0$ appropriate to the parameters of the extrinsic
curvature we are using. We choose therefore a Misner geometry
characterized by the same value of $L$ as in (\ref{Kapprox}). Since
$L$ there represents not the physical distance, in any sense, between
the holes, but the formal distance in the conformally flat space, we
choose a Misner geometry with the same value of $L$ in the conformally
flat part of the Misner metric. The relationship of $L$ to $\mu_0$ is
(see, e.g., \cite{PrPu})
\begin{equation}\label{Lvsmu}
L/M=\frac{\rm{\coth}\mu_0}{2\Sigma_1}\ \ \hspace*{30pt}
  \ \ \ \Sigma_1\equiv\sum_{n=1}\frac{1}{\sinh n\mu_0}\ .
\end{equation}
This completes the description of the initial data to first order in
$L$ and to first order in $P$ (the close, slow approximation).  We now
view the spacetime exterior to the horizon as a perturbation of a
single Schwarzschild hole described in standard Schwarzschild
coordinates $t,r,\theta,\phi$. Even-parity perturbations are then
described by a Zerilli function $\psi$.  According to the general
prescription given in reference \cite{AbPr} the value of $\psi$ on a
$t=0$ initial hypersurface is found from the initial value of the
three geometry. Our initial geometry, to first order in $P$, is exactly
the same as the zero $P$ solution in reference\cite{PrPu}, where the
Zerilli function is denoted $\psi_{\rm pert}$, and is given in eq.
(4.29), along with (4.10),(4.27) and (4.28). In that reference it is
shown that in the close limit, the quadrupole contribution dominates,
with contributions for $\ell>2$ higher order in the separation
parameter. Here we shall consider only the $\ell=2$ contribution, and
shall denote the Zerilli function, corresponding to this Misner (i.e.,
$P=0$)problem, as $\psi_{\rm Mis}(r,t)$.

The initial value of $\dot{\psi}$, the time derivative of the Zerilli
function, follows from the extrinsic curvature as explained in
\cite{AbPr}. The extrinsic curvature is given, in our approximation,
by multiplying $\widehat{K}^{ab}$ in (\ref{Kapprox}) by the squared
reciprocal of the conformal factor for the Schwarzschild geometry,
$\phi_{\rm Schw}=1+M/2R$.
We must map the coordinates of the
initial value solution to the coordinates for the Schwarzschild
background. To do this, we use the same mapping used for the initial
value of $\psi$ in \cite{PrPu}: we interpret the $R$ of
(\ref{Kapprox}) as the isotropic radial coordinate of a Schwarzschild
spacetime, and we relate it to the usual Schwarzschild radial
coordinate $r$ by $R=(\sqrt{r}+\sqrt{r-2M})^2/4$.
{}From this we arrive at the following expression for the (Schwarzschild
coordinate basis) components of the extrinsic curvature:
\begin{equation}\label{finKapp}
K_{ab} =
{3 P L \over 2r^3} \left[
\begin{array}{ccc} \frac{-4 \cos^2 \theta}{1-2M/r}&0&0\\
0& r^2 (1+\cos^2 \theta)&0\\
0&0& r^2 \sin^2 \theta (3 \cos^2 \theta -1)
\end{array}\right]\ .
\end{equation}
Here we have used the fact that
\begin{equation}
\phi^2\approx\phi_{\rm Mis}^2\approx
\phi_{\rm Schw}^2=r/R
=\frac{1}{\sqrt{1-2M/r}}\,\frac{dr}{dR}\ .
\end{equation}
{}From (\ref{finKapp}), which contains both monopole and quadrupole
parts, we can project out the $\ell=2$ part and read off the initial
value of the time derivative of the Zerilli function to be
\begin{equation}\label{initdot}
\dot{\psi}|_{t=0}=-{24 P L \sqrt{1-2M/r}\over r^2 (2+3 M/r)} \sqrt{4 \pi\over
5} (4 + {3 M\over r})\ .
\end{equation}
Along with $\psi|_{t=0}=\psi_{\rm Mis}(r,t=0)$, this completes the
specification of the Cauchy data for $\psi$.

Given this Cauchy data, the time evolution is obtained by evolving the
Zerilli equation,
\begin{equation}\label{zerilli}
\partial^2 \psi / \partial t^2 -
\partial^2 \psi / \partial r_*^2 +V(r) \psi = 0\ ,
\end{equation}
where $r_*=r+\log(r/2M-1)$ and the Zerilli function $\psi$ is a
coordinate invariant combination of the perturbed metric coefficients;
the $\ell=2$ ``potential'' V(r) can be seen in reference \cite{PrPu}.

\begin{figure}
\hspace{4cm} \epsfxsize=200pt \epsfbox{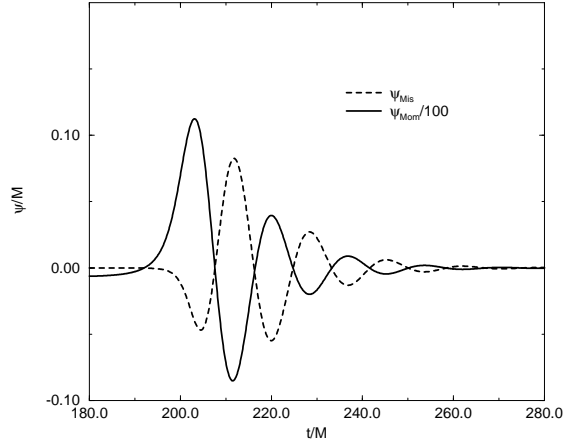}
\caption{The anticorrelation between the
two components of the perturbation,
 $\psi_{\rm Mis}$ and $\psi_{\rm Mom}$,
  leads to the negative coefficient in the energy vs.
  momentum relation.}\label{anticorr}
\end{figure}

\begin{figure}
\hspace{4cm} \epsfxsize=200pt \epsfbox{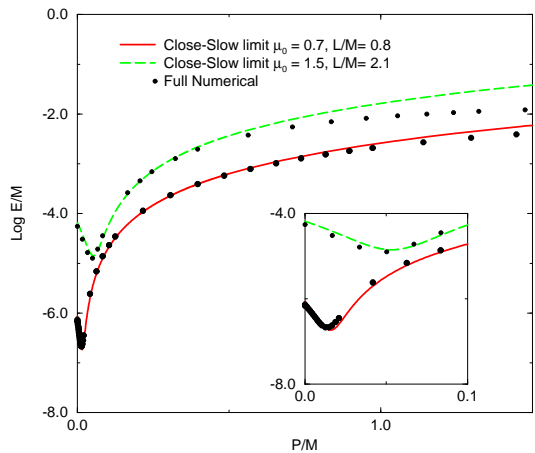}
\caption{Energy as a function of initial momentum. Here $E$ is the
energy radiated
  during coalescence, $P$ is the initial momentum, and $M$ is the
  initial ADM mass. Curves shown are for fixed parameter
  $\mu_0$, corresponding to separation of the holes in conformal
  space. The curves clearly show the ``dip'' effect, and the good
  agreement, even for large values of the momentum.}\label{energy}
\end{figure}

The evolved $\psi$ can be decomposed into two
components
\begin{equation}\label{decomp}
\psi=\psi_{\rm Mis}+\psi_{\rm Mom}\ .
\end{equation}
The first term is the solution of (\ref{zerilli}) for cauchy data
$\psi=\psi_{\rm Mis}(r,t=0)$ and $\dot{\psi}=0$ at $t=0$. The second
term is the solution for $t=0$ cauchy data $\psi=0$, and with
$\dot{\psi}$ given by (\ref{initdot}). The two contributions are
respectively zero order in $P$ and first order in $P$; the
decomposition then represents a separation into parts of $\psi$ due to
the masses, and to the momenta.
The radiated energy is given by\cite{Anninosetal}
\begin{equation}
E=  {1 \over 384 \pi}\int_0^\infty
 \dot{\psi}^2 dt \ ,
\end{equation}
and can be written, in terms of the decomposition above, as:
\begin{eqnarray}\label{energyint}
E =  {1 \over 384 \pi} \left(
\int_0^\infty\dot{\psi}_{\rm Mis}^2\ dt
+2\int_0^\infty\dot{\psi}_{\rm Mis}\dot{\psi}_{\rm Mom}\ dt
+\int_0^\infty\dot{\psi}_{\rm Mom}^2\ dt\right)\label{erad}\ .
\end{eqnarray}
The first term gives the same result as in the momentarily stationary
case; it is simply the radiation for the Misner initial geometry, as
computed in reference\cite{PrPu}. The second term is linear in the
momentum of each hole. The coefficient of it is given by the
``correlation'' of $\psi_{\rm Mis}$ and $\psi_{\rm Mom}$. As can be
seen in Fig.~\ref{anticorr}, this ``correlation'' integral is
negative.
The anticorrelation is compatible with previous simulations done by
Ref. \cite{ac94}
using numerical initial data (see figures 3a,b in
their paper).  This means that for small values of $P$, the radiated
energy {\em decreases} with increasing momentum. The effect is clearly
visible in Fig.~\ref{energy} where we show the radiated energy as a
function of the momentum,

Note that the first term is simply a function of the Misner
parameter $\mu_0$.  The second term depends on $\mu_0$, but also
depends on $L$ and $P$.  We can write $L$ in terms of $\mu_0$ with
(\ref{Lvsmu}) to express all  dependencies in (\ref{energyint}) only in terms
of $\mu_0$ and $P/M$. With the correct numerical factors we get the
final result of the close-slow approximation, a simple formula for
the radiated energy simply and explicitly expressed in terms of the parameters
of the collision:
\begin{equation}\label{clsleq}
\frac{E}{M}=2.51\times10^{-2}\kappa_2^2(\mu_0)
-2.06\times10^{-2}\frac{{\rm \coth}\,\mu_0\kappa_2(\mu_0)}{\Sigma_1}\left(
\frac{P}{M}
\right)
+5.37\times10^{-3}\left(
\frac{{\rm \coth}\,\mu_0}{\Sigma_1}
\right)^2
\left(
\frac{P}{M}
\right)^2\ ,
\end{equation}
where $\kappa_2$, as defined in Ref.~\cite{Anninosetal}, is
\begin{equation}\label{kapdef}
\kappa_2(\mu_0)\equiv\frac{1}{\left(4\Sigma_1\right)^3}
\sum_{n=1}^\infty
\frac{(\coth{n\mu_0})^2}{\sinh{n\mu_0}}\ .
\end{equation}
In Fig. \ref{energy}, we plot the radiated energy computed from
(\ref{clsleq}) for several values of initial separation $\mu_0$, and
for a wide range of $P/M$ . On this plot, also, are presented the
results for radiated energy from numerical results computations which
make no approximations.  The agreement between the numerical results and the
results of the approximation is remarkably good, even at
rather large values of $P/M$.

\section{Results}

Two features of Fig.~\ref{energy} stand out. The first is ``momentum
dominance'': the radiated energy is dominated by the third integral in
(\ref{energyint}) unless the momentum is very small. The second
obvious feature is that the approximation method works very well even
for sizeable values of $P/M$.

To understand the implications of these features, let us start by
reviewing the difference between the exact, nonlinear numerical
computation, and the approximation scheme of Sec.~III.  In the exact
method we start with an exact solution to the initial value equations
described by two parameters, one a dimensionless measure of the
separation of the holes, the other a dimensionless measure of the
momentum. The process of generating the solution consists of four
steps: (i) One starts with a very simple prescription for
$\widehat{K}_{ab}$ constructed by superposing two solutions of form
(\ref{onehole}) corresponding to two coordinate positions in the
conformally flat space.  (ii) Equation (\ref{hamil}) is then solved
for the conformal factor and hence for the three geometry.  (iii) The
solution for the extrinsic curvature and the initial geometry is then
``symmetrized'' by an iterative process equivalent to adding image
charges. (iv) This solution is numerically evolved off the initial
hypersurface with the full nonlinear Einstein equations.

By contrast, the steps for the approximate solution are: (i) The
(conformal) extrinsic curvature is taken to be the unsymmetrized
superposition of two contributions with the form of
(\ref{onehole}). (ii) The conformal factor, and therefore the three
geometry, is taken to be the symmetrized solution corresponding to
throats located at the same points in the conformally flat space as
the points in $\widehat{K}_{ab}$. (iii) This approximate initial data
is then treated as initial data for the nonspherical perturbations of
a Schwarzschild hole, and the perturbations are evolved with the
linearized Einstein equations.

The difference in evolution off the initial hypersurface (full
Einstein equations in one case, linearized equations in the other) is
not a major source of error in the interesting cases, those with high
momentum.  As momentum increases, the location of the horizon in the
initial geometry moves outward.  The high momentum cases, therefore,
correspond to throats which, on the initial hypersurface, are well
inside an all-encompassing horizon.  This is the situation in which
the ``close-limit'' approximation method should work very well.  It is
also not surprising that no large error is introduced by the failure,
in the approximation method, to symmetrize the extrinsic curvature.
One way of understanding this is to note that $\psi_{\rm Mom}$ lacks
the ``image'' contributions needed for symmetrization. These images
only influence the form of $\widehat{K}_{ab}$ very close to the
holes. As the separation between the holes gets smaller the horizon
moves further from the throats and the effect of the images on
$\widehat{K}_{ab}$ outside the horizon diminishes.  We have checked
numerically that the difference between the symmetrized and
unsymmetrized $\widehat{K}_{ab}$, for all cases considered, is
negligibly small outside the horizon.

These two aspects of the approximation method rely on the throats
being ``close'' in some sense, an approximation that seems well
justified. What remains to be explained is how the slow-limit
approximation does such a good job of approximating the very
``unslow'' correct initial data.  We must also justify the apparent
inconsistency in how the approximation scheme deals with orders of
$P$. In the computation of $\psi$ the scheme explicitly omits
corrections of order $P^2$ in (\ref{hamil}).  Formally, then, we
should only be able to keep terms of first order in $P$ in
(\ref{energyint}).  But it is the apparently inconsistent $P^2$ terms,
of course, which dominate at most points in Fig.~\ref{energy}
(``momentum dominance'' in generation of radiation). Not
only do the $P^2$ terms agree with the results of numerical
relativity, but the agreement remains good for rather high values of
$P/M$. This raises the question: just what momentum contributions has
our approximation really omitted?

The momentum enters into the construction of the initial data in only
two direct ways. First, it is an overall scaling parameter for
$\widehat{K}_{ab}$.  The expression in (\ref{Kapprox}) is an
approximation for small $L$, but it is exact in $P$.  The process of
symmetrizing does not change this. Up to a conformal factor, then, the
extrinsic curvature is exactly linear in $P$. Second, $P$ enters the
determination of the conformal factor through (\ref{hamil}).  The
success of the slow approximation must be directly ascribed to the
relatively unimportant role played by the right hand side of
(\ref{hamil}).

Further work will be needed for a real understanding of this, but some
reasonable speculations can already be made.  Due to momentum
dominance the details of the initial three geometry are not crucial,
so any quadrupolar distortion induced by $\widehat{K}_{ab}$ at large
$P$ will be insignificant compared to the radiation generated by the
extrinsic curvature. The ``slow'' approximation, of course, is not
perfect; at sufficiently high momentum it begins to fail. We speculate
that the reason for this failure is not primarily due to
$\widehat{K}_{ab}$ generating quadrupolar distortions of the initial
three geometry. Rather, it is the effect of that source on the
monopole part of the conformal factor, and hence on the ADM mass
``$M$,'' that is used to scale physical quantities. When we do a
comparison in Fig.~\ref{energy} between the numerical relativity
results and those of the approximation, we are comparing two cases for
the same $\mu_0$ (i.e.,  the same coordinate separation in conformal
space) and for the two cases we compare $E/M$ at a given value of
$P/M$. We are therefore placing on an equal footing the true value of
$M$ in the numerical relativity solution, and the $P\rightarrow0$
value of $M$ in the approximation. It should be possible, in
principle, to correct for this and, in effect, reduce the
approximation to one in which we have only ignored the quadrupolar
part of the source in (\ref{hamil}).

The present results greatly help us to understand the success of the
results of Ref. \cite{ac94}. That
\begin{figure}
\hspace{4cm} \epsfxsize=200pt \epsfbox{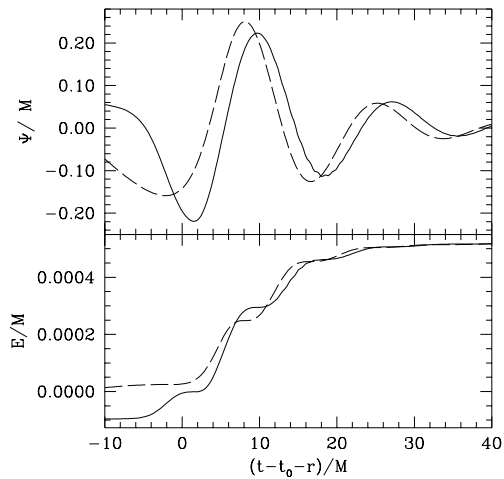}
\caption{
Radiated energy as a function of time for two different initial
value sets.  The first is for equal mass holes falling from rest
at $\mu_0=2.2$.  The second shows the result of a boosted collision
starting from a separation parameter $\mu_0=1.406$ and a momentum
$P/M = 0.23$.  The second set of initial data can be
considered to be an approximation to a constant $t$ slice
of the spacetime that evolves from the first set.
The time scale is the (flat space) retarded time with
zero corresponding to the time of apparent horizon formation.
The energy scale of the $\mu_0=2.2$ curve has been offset to
zero at the time of apparent horizon formation.}\label{time}
\end{figure}
success seems to require two things about the generation of
gravitational radiation in collisions from large distances: (i) There
must be negligible radiation during the early motion, when the holes
are in each other's weak field region.  (ii) The only important
consequence of the early, weak-field, motion
must be to give the
holes momentum when each reaches the strong field region of the other.
The first requirement is relatively easy to check. In Fig.~\ref{time}
we plot radiated energy, computed by methods of numerical relativity,
as a function of time, first for initial data representing two black
holes falling from large separation. (The oscillations are due to the
fact that almost all the energy comes off as ``quasinormal ringing''
of the final hole formed.) We also show the result of a second
calculation.  Cook\cite{cook91} initial data are taken corresponding
to the separation and momentum that the black holes would have after
falling to a fairly close separation. A comparison of the curves
verifies that the early stage of motion does not produce a significant
contribution to the total outgoing radiation.

Our present results, and in particular momentum dominance, strongly
support the second requirement for the success of the ideas of
Ref. \cite{ac94}. Since
$\psi_{\rm Mom}$ is the source of essentially all the radiation,
one can see that what is important about the early stages of the
coalescence is only the development of extrinsic curvature.  This does
not, of course, explain why there seems to be insensitivity to the
details of the extrinsic curvature. (Surely, the Bowen-York extrinsic
curvature, symmetrized or not, is not actually the extrinsic curvature
that evolves from earlier stationary conditions. Yet, it seems to be
adequate to give good predictions.)
  A more satisfactory answer
to this question means that we must understand the relationship
between data on an initial hypersurface and how this evolves to data
on subsequent hypersurfaces. We must also understand the importance of
confining ourselves to conformally flat data on
hypersurfaces. Progress on these questions will probably require
comparable results from four distinct classes of initial data sets.
These are (a) Misner data with large hole separation, (b) the non
conformally-flat data with close holes that evolves from (a), (c)
boosted conformally-flat data with close holes, and (d) boosted
conformally-flat data in the close-slow approximation.  In addition,
one requires reasonable measures of physical separation and momentum
so that correspondence can be drawn between disparate initial data
sets.

There is strong motivation for carrying out such studies.  The results
so far achieved, both by numerical relativity and with the close and
the slow approximation, are limited to head-on collisions. The
situation of astrophysical interest, of course, is very different: the
coalescence of orbiting holes. If the last few orbits in a coalescence
are to be studied with numerical relativity, it will be crucial to
understand what initial data are to be used to start the computation.
Studies with the head-on collision provide a useful starting point
to understanding the sensitivity of the radiation generation to the details
of the initial data.

A rather different, and more speculative, motivation for a better
understanding of these issues, is the hope that our approximation
methods might be as successful with orbital problems as with head-on
coalescence. These results might provide ``easy'' approximate answers
over a reasonable range of orbital coalescences, and may therefore
serve as a guide to the numerical studies.

\acknowledgments

This work was supported in part by grants NSF-PHY-9423950,
NSF-PHY-9396246, NSF-PHY-9207225, NSF-PHY-9507719, NSF-PHY-9407882,
research funds of the Pennsylvania State University, the University of
Utah, the Eberly Family research fund at PSU and PSU's Office for
Minority Faculty development. JP acknowledges support of the Alfred
P. Sloan foundation through a fellowship.

\end{document}